\documentclass[twocolumn,amsmath,amssymb,prl]{revtex4}
\usepackage{latexsym}
\usepackage{graphicx}
\usepackage{color}
\begin{document}
\title{Gating of high-mobility InAs metamorphic heterostructures}

\author{J.~Shabani$^{1}$}
\author{A.~P.~McFadden$^{2}$}
\author{B.~Shojaei$^{3}$}
\author{C.~J.~Palmstr{\o}m$^{1,2,3}$}
\affiliation{$^{1}$California NanoSystems Institute, University of California, Santa Barbara, CA 93106, USA
\\
$^{2}$Department of Electrical Engineering, University of California, Santa Barbara, CA 93106, USA
\\
$^{3}$Materials Department, University of California, Santa Barbara, CA 93106, USA
}
\date{\today}
\begin{abstract}
We investigate the performance of gate-defined devices fabricated on high mobility InAs metamorphic heterostructures. We find that heterostructures capped with In$_{0.75}$Ga$_{0.25}$As often show signs of parallel conduction due to proximity of their surface Fermi level to the conduction band minimum. Here, we  introduce a technique that can be used to estimate the density of this surface charge that involves cool-downs from room temperature under gate bias. We have been able to remove the parallel conduction under high positive bias, but achieving full depletion has proven difficult. We find that by using In$_{0.75}$Al$_{0.25}$As as the barrier without an In$_{0.75}$Ga$_{0.25}$As capping, a drastic reduction in parallel conduction can be achieved. Our studies show that this does not change the transport properties of the quantum well significantly. We achieved full depletion in InAlAs capped heterostructures with non-hysteretic gating response suitable for fabrication of gate-defined mesoscopic devices.

\end{abstract}
\maketitle
Narrow band gap semiconductors such as InAs are of fundamental interest for next generation high-speed electronics due to their unique material properties of small effective mass, large dielectric constant and high room temperature mobility \cite{Kroemer04, AlamoNature11}. In addition, they possess strong spin orbit interaction and large g-factor which make them an ideal platform for spintronics applications \cite{WolfScience01,ZuticRevModPhys04}. Recently, two-dimensional electron systems (2DESs) confined to InAs layers have become the focus of renewed theoretical and experimental attention partly because of their potential applications in quantum computation \cite{Alicea10,QCpaper, NadjPergeNature10}. However, all these applications require precise control of electrostatic potentials and carrier densities using nano-fabricated metallic gates. Unlike widely used GaAs based systems, reliable gating has proven difficult in InAs based systems due to gate leakage and hysteretic behavior. Charge traps and surface Fermi level pinning could drastically affect the device performance. Controlling surface properties in In$_{x}$Ga$_{1-x}$As material has played a crucial role in achieving high quality interfaces for metal oxide semiconductor field effect transistors (MOSFETs) \cite{AlamoNature11}. 

\begin{figure}[tp]
\centering
\includegraphics[scale=0.45]{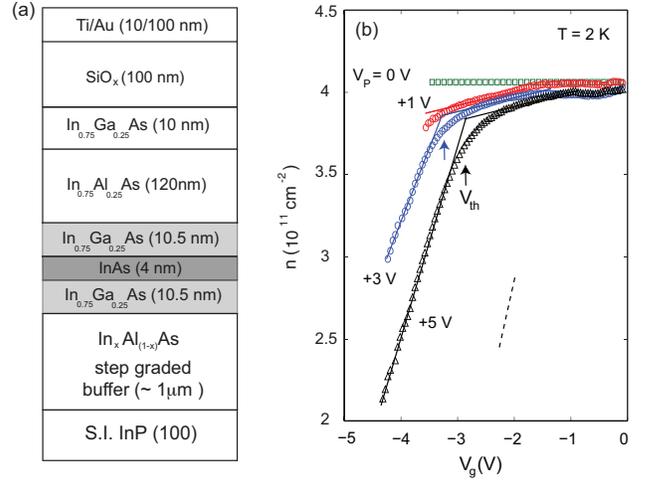}
\caption{(Color online) (a) Schematic of our InAs heterostructure material stack. Gated Hall bars are fabricated using PECVD grown SiO$_{x}$ dielectric and Ti/Au metallic gates. (b) Electron density, $n$ as a function of top gate bias, $V_{g}$ for different cool-down biases, V$_{P}$ = 0 V, +1 V, +3 V, +5 V. There are two slopes in the data as the gate bias is swept negative; first the surface charge is depleted and then after a threshold voltage, V$_{th}$, electrons are removed from the quantum well. The threshold voltage for each V$_{P}$ is extrapolated from these two slopes, shown by arrows for $V_{P}$= +3 V and +5 V. The dashed line shows the expected slope when the surface charge is depleted.}
\end{figure}

The Fermi level pinning at semiconductor surfaces has been the subject of numerous theoretical and experimental studies \cite{SzeBook}. In most semiconductors, such as GaAs, the Fermi level is pinned in the band gap \cite{MeadPR64}. It is well known that the surface states in case of InAs (100) can result in a two dimensional electron system \cite{TsuiPRL70}. The electron accumulation is due to pinning of the Fermi level above the conduction band minimum. The position of the pinning level sensitively depends on the material and the surface treatments \cite{Sulfur}. In the case of high indium content InGaAs, the situation is similar to InAs. Experiments on In$_{x}$Ga$_{1-x}$As predicts Schottky barrier height becomes negative, exhibiting an ohmic behavior, for $x > 0.85$ \cite{KajiyamaAPL73}. In this work, we grow InAs metamorphic heterostructures with an In$_{0.75}$Ga$_{0.25}$As surface layer, see Fig.~1(a). High indium content InGaAs, in this case $x=0.75$, is required for growth of a strained InAs layer \cite{InAscriticalthickness} and to minimize dislocations in the active region, InAlAs barriers and surface layers are all grown at this indium composition. The surface Fermi level pinning in In$_{0.75}$Ga$_{0.25}$As layers is theoretically estimated to be 40 meV below the conduction band minimum. However, we often find signatures of parallel conduction in heterostructures similar to Fig.~1(a). In this paper, we investigate the transport properties of these heterostructures and show that in this type of heterostructure, In$_{0.75}$Al$_{0.25}$As surface layers are better suited for gate-defined devices.

The samples were grown on a semi-insulating InP (100) substrate, using a modified VG-V80H molecular beam epitaxy system. After oxide desorption under an As$_{4}$ overpressure at 520 $^{o}$C, the substrate temperature is lowered to 480 $^{o}$C where we grow a superlattice of InGaAs/InAlAs lattice matched to InP. The substrate temperature is further lowered to 360 $^{o}$C for growth of the In$_{x}$Al$_{1-x}$As buffer layer. This lower temperature buffer growth is used to reduce and minimize the influence of dislocations forming due to the lattice mismatch of the active region to the InP substrate. The indium content in In$_{x}$Al$_{1-x}$As is step graded from $x=$0.52 to 0.85 \cite{Richter00,Wallart05}. We lower the indium content to x=0.75 while increasing the substrate temperature to $T_{sub} \sim$ 490 $^{o}$C with As$_{4}$ beam equivalent pressure (BEP) to $2.0 \times 10^{-5}$. The quantum well consists of a 4 nm InAs layer sandwiched by In$_{0.75}$Ga$_{0.25}$As layers for both samples.  An In$_{x}$Al$_{1-x}$As (120 nm) top barrier layer is grown next as shown in Fig.~1(a). Finally, the structures are capped with a 10 nm In$_{0.75}$Ga$_{0.25}$As layer.  The structures are nominally undoped but the deep levels in InAlAs donate carriers to the quantum well \cite{Capotondi04}.

\begin{figure}[tp]
\centering
\includegraphics[scale=0.5]{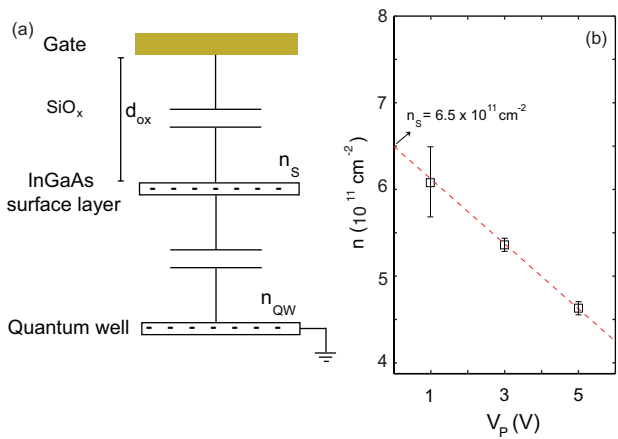}
\caption{(Color online)  (a) Schematic of capacitive model of our system. (b) Extracted surface density as a function of cool-down bias. The fit to the data is shown red dashed line and estimates a surface charge of $n_{S}=6.5 \times 10^{11}$ cm$^{-2}$ at zero bias cool-down.}
\end{figure}

\begin{figure}[t]
\centering
\includegraphics[scale=0.45]{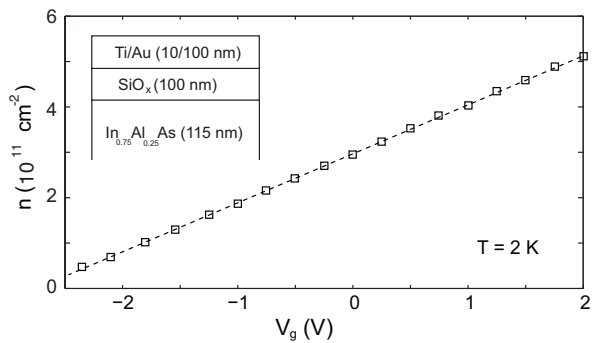}
\caption{(a) Density as a function of top gate bias under no cool-down bias. The density can be tuned over a wide range from $5 \times 10^{11}$ cm$^{-2}$ down to near depletion. The material stack is shown near surface. This sample is taken from the same wafer as shown in Fig.~1. The only different is that InGaAs surface layer is removed using wet etch. }
\end{figure}

We fabricated gated-Hall bars by deposition of 100 nm SiO$_{x}$ at a substrate temperature of 250 $^{o}$C using plasma enhanced chemical vapor deposition (PECVD) and a subsequent 10/100 nm Ti/Au as a metallic gate. The Hall bars are cooled down to 2 K with no bias applied on the top gate, $V_{P}$ = 0 V. At $T$ = 2 K, magneto-resistance exhibits Shubnikov de Haas oscillations and well defined integer quantum Hall states in strong magnetic field \cite{ShabaniMIT}. The typical electron mobility is measured to be $\mu \sim 2 \times 10^{5}$ cm$^{2}$/Vs at $n \sim 4 \times 10^{11}$ cm$^{-2}$. We present data on one such Hall bar but they all show qualitatively similar data. The carrier densities are determined using Hall data for each gate bias, $V_{g}$. The green squares in Fig.~1(b) show no change in carrier density of the quantum well as a function of $V_{g}$. This behavior suggests formation of surface charge that shields the electric field from reaching the quantum well. 

Application of a positive bias, $V_{P}$, at room temperature and during cool-down changes the gating response. Figure 1(b) shows electron density dependence as a function of $V_{g}$ for several positive cool-down biases, $V_{P}$ = +1 (red), +3V (blue) and +5V (black). This suggests that the surface charge density is lowered with an applied gate bias during cool-down. Positive bias cool-down has been used in GaAs material system for improving noise response of mesoscopic devices  \cite{CharlesMarcus}. It is believed that {\it parasitic} electrons can be frozen in donor sites (DX centers) during cool-down. Similarly, positive bias during cool-down could freeze the  electrons to their donor sites (for example dangling bonds) and hence there will be less free surface charge after cool-down under bias. The presence of a threshold gate voltage before the 2DES electron density can be tuned even at our highest positive bias cool-down, $V_{P}$= +5 V, suggests there is still some remaining surface charge. There are two slopes in the data as the gate bias is swept negative; first the electrons are depleted from the InGaAs surface states but due to finite density of states in the surface layer, there is a slight slope in the density of the quantum well \cite{Sluryi, JimAPL91}. After all the electrons from the surface states are depleted (at the threshold voltage), we observe the electron density in the quantum well is decreasing with a slope that closely matches the calculated slope (dashed line in the inset) using the total capacitance from the top gate to the quantum well, see Fig.~2(a).  The threshold voltage for each V$_{P}$ is extrapolated from these two slopes, shown by arrows for $V_{P}$= +3 V and +5 V. Using our cool-down data, we can estimate the surface charge by converting the threshold voltages into charge density using $n_{S} = (\epsilon/d_{ox}) \times V_{th}/e$. These values are plotted in Fig.~2(b). A linear fit through the data estimates $n_{S} = 6.5 \times 10^{11}$ cm$^{-2}$ at the surface when no bias is applied, $V_{P} = 0$.

\begin{figure}[t]
\centering
\includegraphics[scale=0.6]{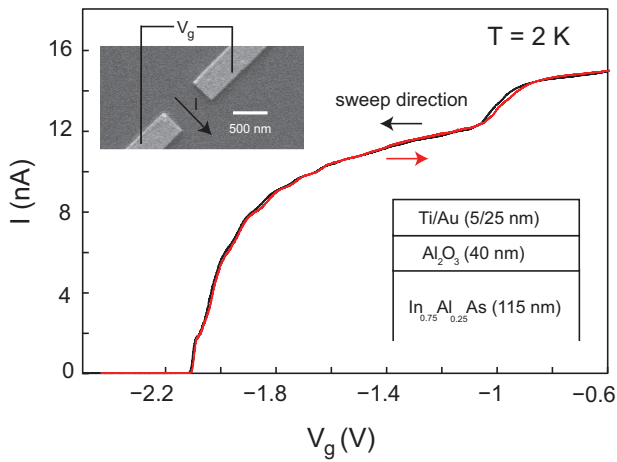}
\caption{(Color online)  Plot of current as a function of side-gate bias shows full pinch off with features related to quantum confinement in 1D. Inset: SEM image of the quantum point contact and material stack. }
\end{figure}

When the InGaAs capping layer is removed using wet chemical etching (H$_{3}$PO$_{4}$$:$H$_{2}$O$_{2}$$:$H$_{2}$O, 1:1:80), we do not observe a drastic change in the quality of the magneto-transport of the 2DES, however, the gating properties changes dramatically. Figure 3 shows the gate bias dependence of the denstiy for a InAlAs capped device. Even when the device is cooled down under no bias, the density could be varied in a wide range from $5 \times 10^{11}$ cm$^{-2}$ down to near depletion. The InAlAs has a higher band gap ($\sim$ 0.9 eV) and the surface Fermi level is expected to be near midgap. The density of the quantum well drops as much as 25\% after removing the InGaAs surface layer. The measured mobility as a function of density at 2 K is well-fitted by $\mu \sim n^{\alpha}$, with $\alpha = 0.8$ at high density range ($n > 2 \times 10^{11}$ cm$^{-2}$). A value of $\alpha \sim 0.8$ signifies that the mobility is limited by scattering from nearby background charged impurities~\cite{DasSarmaPRB13, ShabaniMIT}. Fabricated quantum devices using these heterostructures such as quantum point contacts (QPCs) show no hysteretic behavior under different sweep directions or by time. Figure 4(b) shows the plot of current through a QPC measured as a function of gate voltage applied to a split-gate device. We achieve full depletion (I $<$ 1 pA, limited by our instrumentation) and observe features associated with quantum confinement in 1D. 

In conclusion, we have studied the gating response of high mobility InAs metamorphic heterostructures. We find that the surface of InGaAs often shows signs of parallel conduction and limits the gating response. Using cool-downs with positive bias we reduced the electron density of these parallel conduction and estimated their density. Further, InAlAs surface layers could be used in these heterostructures and our studies show that not only the quality of the quantum well is maintained but also they show full depletion and non-hysteretic gating response due to drastic reduction of surface charge.

Our work was supported by the Microsoft Research. A portion of this work was performed in the UCSB nanofabrication facility supported by the NSF through the National Nanotechnology Infrastructure Network.

\begin{center}
{\bf References}
\end{center}

\begin{thebibliography}{99}
\bibitem{InAscriticalthickness}
The critical thickness of growing InAs on In$_{0.75}$Ga$_{0.25}$As is about 6 nm.
\end{thebibliography}
\end{document}